\documentclass[prd,preprint,superscriptaddress,showpacs,byrevtex]{revtex4}
\usepackage{bm}
\def\fsl#1{\setbox0=\hbox{$#1$}           
   \dimen0=\wd0                                 
   \setbox1=\hbox{/} \dimen1=\wd1               
   \ifdim\dimen0>\dimen1                        
      \rlap{\hbox to \dimen0{\hfil/\hfil}}      
      #1                                        
   \else                                        
      \rlap{\hbox to \dimen1{\hfil$#1$\hfil}}   
      /                                         
   \fi}                                         %
\newcommand{\be}{\begin{equation}}
\newcommand{\ee}{\end{equation}}
\newcommand{\bea}{\begin{eqnarray}}
\newcommand{\eea}{\end{eqnarray}}
\newcommand{\beq}{\begin{equation}}
\newcommand{\eeq}{\end{equation}}
\newcommand{\beqs}{\begin{eqnarray}}
\newcommand{\eeqs}{\end{eqnarray}}

\newcommand{\aslash}{A\hspace{-0.067in}\slash}

\begin{document}
\title{Parton To Hadron Fragmentation Function From Quark-Gluon Plasma Using Lattice QCD Method At Finite Temperature }
\author{Gouranga C Nayak }\thanks{E-Mail: nayakg138@gmail.com}
%
%
\date{\today}
\begin{abstract}
Recently we have shown that it is possible to study the parton to hadron fragmentation function by using the lattice QCD method at the zero temperature. In this paper we extend this to the finite temperature QCD and study the parton to hadron fragmentation function from the quark-gluon plasma by using the lattice QCD method at the finite temperature.
\end{abstract}
\maketitle
\pagestyle{plain}

\pagenumbering{arabic}

\section{ Introduction }

Just after $\sim 10^{-12}$ seconds of the big bang the universe was filled with a hot and dense state of the matter known as the quark-gluon plasma (QGP). The temperature of the quark-gluon plasma is $\sim 10^{12}$K which is about a million times larger than the temperature of the sun. This temperature corresponds to the energy density $\sim$ 2 GeV/fm$^3$ which is higher than the energy density 0.15 GeV/fm$^3$ of the nucleus. Similarly, after the black hole, the quark-gluon plasma is the densest state of the matter in the universe. Hence it is important to recreate this early universe scenario in the laboratory, {\it i. e.}, to create the quark-gluon plasma in the laboratory.

There are two experiments which produce the quark-gluon plasma in the laboratory; 1) the relativistic heavy-ion colliders (RHIC) at BNL where two gold nuclei collide at the total center of mass energy equals to 200$\times $197 GeV, and 2) the large hadron colliders (LHC) at CERN where two lead nuclei collide at the total center of mass energy equals to 5.02$\times $208 TeV \cite{qp1,qp2,qp3}. Since these huge energies are deposited over small volumes (just after the nuclear collisions) there is no doubt that the energy density $\sim$ 2 GeV/fm$^3$ to produce the quark-gluon plasma is reached at RHIC and LHC heavy-ion colliders.

It is well known that the quantum chromodynamics (QCD) is the fundamental theory of the nature describing the interaction between quarks and gluons \cite{ymp}. Due to the asymptotic freedom the renormalized QCD coupling becomes weaker at the short distance \cite{gwp} where the perturbative QCD (pQCD) is applicable. Hence at the high energy colliders the short distance partonic scattering cross section is calculated by using the pQCD. However, this partonic cross section cannot be directly experimentally measured because we have not directly experimentally observed quarks and gluons. By using the factorization theorem in QCD \cite{fcp,fc2p} this partonic scattering cross section is convoluted with the parton distribution function (PDF) and with the fragmentation function (FF) to calculate the hadron production cross section at the high energy colliders which is experimentally measured.

Because of this reason the parton to hadron fragmentation function (FF) plays a key role at the high energy colliders. Note that the PDF and FF are extracted from the experiments because the analytical solution of the non-perturbative QCD is not known yet, {\it i. e.}, the path integration in QCD cannot be done analytically due to the presence of the cubic and quartic gluonic field terms in the QCD lagrangian density (see section II for details).

For this reason the path integration in QCD is done numerically in the Euclidean time by using the lattice QCD method. Hence the lattice QCD method can study the hadron formation from the quarks and gluons \cite{nkhp}. Similarly the lattice QCD method can be employed to calculate the parton distribution function (PDF) inside the hadron \cite{nkpp}.

However, in the literature it is claimed that the parton to hadron fragmentation function (FF) cannot be studied by using the lattice QCD method because of the sum over the outgoing inclusive (unobserved) hadrons. But recently we have reported that since the hadron formation from the quarks and gluons can be studied in the lattice QCD method \cite{nkhp}, the parton to hadron fragmentation function (FF) can be studied in the lattice QCD method by using the LSZ reduction formula at the partonic level \cite{nkfp}.

As mentioned above since the quarks and gluons are not directly experimentally observed we cannot directly detect the quark-gluon plasma at RHIC and LHC. Due to this reason the indirect signatures are proposed for the quark-gluon plasma detection. The indirect hadronic signatures for the quark-gluon plasma detection at the RHIC and LHC heavy-ion colliders are: 1) jet quenching, 2) $j/\psi$ suppression, and 3) strangeness enhancement.

Since we do not directly detect the quark-gluon plasma but we directly experimentally measure the hadrons at RHIC and LHC it is necessary to understand how the quarks and gluons from the quark-gluon plasma fragment to hadrons which are experimentally measured. Hence it is necessary to study the parton to hadron fragmentation function from the quark-gluon plasma from the first principle. The numerical lattice QCD method at the finite temperature QCD in the Euclidean time formulation is the first principle method to study the quark-gluon plasma at the finite temperature in the quantum field theory.

As discussed above we have recently studied the parton to hadron fragmentation function (FF) in the lattice QCD method at the zero temperature by using the LSZ reduction formula at the partonic level \cite{nkfp}. In this paper we extend this to the finite temperature QCD and study the parton to hadron fragmentation function (FF) from the quark-gluon plasma by using the lattice QCD method at the finite temperature.

The paper is organized as follows. In section II we discuss the parton to hadron fragmentation function (FF) by using the lattice QCD method at the zero temperature. In section III we study the partonic process at the finite temperature which is necessary to study the parton to hadron fragmentation function from quark-gluon plasma. In section IV we study the parton to hadron fragmentation function (FF) from the quark-gluon plasma by using the lattice QCD method at the finite temperature. Section V contains conclusions.

\section{ Parton to hadron fragmentation function by using the lattice QCD method at the zero temperature }

In order to study the hadron formation from quarks and gluons using lattice QCD method one chooses the partonic operator ${\cal B}^H(x)$ such that it contains the same quantum numbers of the hadron $H$. In this paper we will consider the light quark $q$ in the quark-gluon plasma at the finite temperature $T$ to fragment to the heavy quarkonium $\eta^c$ [the $^1S_0$ bound state hadron of the charm quark-antiquark pair $c{\bar c}$], {\it i. e.}, we will study the fragmentation function $D_{q \rightarrow \eta^c}$ for the process
\bea
q \rightarrow \eta^c+X
\label{etcp}
\eea
where $X$ represents the inclusive (unobserved) particle(s). The extension of this procedure to other partons and to other hadrons is straightforward.

The partonic operator ${\cal B}^{\eta_c}(x)$ to form $\eta^c$ from $c{\bar c}$ is given by
\bea
{\cal B}^{\eta_c}(x)={\bar \psi}_i^c(x)\psi_i^c(x)
\label{etp}
\eea
where $\psi_i^c(x)$ is the charm quark field with color index $i=1,2,3$. In QCD in vacuum ({\it i. e.}, without quark-gluon plasma) the non-perturbative correlation function $<0|{\cal B}^{\eta_c}(t,r) {\cal B}^{\eta_c}(0)|0>$ of the partonic operator ${\cal B}^{\eta_c}(x)$ is given by
\bea
&& <0|{\cal B}^{\eta_c}(t'',r'') {\cal B}^{\eta_c}(0)|0>=\frac{1}{Z[0]} \int [dA] [d{\bar \psi}^c][d\psi^c]\Pi_{f=1}^3 [d{\bar \psi}^f] [d\psi^f]\times {\cal B}^{\eta_c}(t'',r'') {\cal B}^{\eta_c}(0) \times {\rm Det}[\frac{\delta H^d}{\delta \omega^c}] \nonumber \\
&& \times {\rm exp}[i \int d^4x [-\frac{1}{4} F_{\mu \delta}^h(x) F^{\mu \delta h}(x) -\frac{1}{2\alpha} [H^d(x)]^2 +\sum_{f=1}^3 {\bar \psi}_j^f(x) [\delta^{jn}(i{\not \partial}-m_f)+gT^h_{jn}\aslash^h(x)]\psi_n^f(x) \nonumber \\
&&+ {\bar \psi}_j^c(x) [\delta^{jn}(i{\not \partial}-M)+gT^h_{jn}\aslash^h(x)]\psi_n^c(x)
\label{npcp}
\eea
where $|0>$ is the full QCD vacuum state or the non-perturbative QCD vacuum state (not the pQCD vacuum state), $\alpha$ is the gauge fixing parameter, $H^d(x)$ is the gauge fixing term [which in covariant gauge is given by $H^d(x) =\partial^\lambda A_\lambda^d(x)$] and $\psi_i^f(x)$ is the light quark field of flavor $f=1,2,3$=up, down, strange quarks. In eq. (\ref{npcp}) the $Z[0]$ is the generating functional in QCD given by
\bea
&& Z[0]= \int [dA] [d{\bar \psi}^c][d\psi^c]\Pi_{f=1}^3 [d{\bar \psi}^f] [d\psi^f] \times {\rm Det}[\frac{\delta H^d}{\delta \omega^c}] \times {\rm exp}[i \int d^4x [-\frac{1}{4} F_{\mu \delta}^h(x) F^{\mu \delta h}(x) -\frac{1}{2\alpha} [H^d(x)]^2 \nonumber \\
&&+\sum_{f=1}^3 {\bar \psi}_j^f(x) [\delta^{jn}(i{\not \partial}-m_f)+gT^h_{jn}\aslash^h(x)]\psi_n^f(x) + {\bar \psi}_j^c(x) [\delta^{jn}(i{\not \partial}-M)+gT^h_{jn}\aslash^h(x)]\psi_n^c(x)
\label{z0p}
\eea
where
\bea
F_{\mu \delta}^h(x) =\partial_\mu A_\delta^h(x) - \partial_\delta A_\mu^h(x) + gf^{hsa} A_\mu^s(x) A_\delta^a(x).
\label{fmp}
\eea
Note that in eqs. (\ref{npcp}) and (\ref{z0p}) we do not have any ghost fields because we directly work with the ghost determinant ${\rm Det}[\frac{\delta H^d}{\delta \omega^c}]$.

The color singlet heavy quarkonium $\eta^c$ formation from the charm quark and anticharm quark operator ${\cal B}^{\eta_c}(x)$ in eq. (\ref{etp}) can be studied from the non-perturbative correlation function in eq. (\ref{npcp}) by using the lattice QCD method at the zero temperature \cite{nkhp}. Since the $\eta^c$ formation from the quarks and gluons can be studied by using the lattice QCD method one finds that the parton to $\eta^c$ fragmentation function can be studied by using the lattice QCD method.

One finds that the $q \rightarrow \eta^c$ fragmentation function $D_{q \rightarrow \eta^c}$ using the lattice QCD method at the zero temperature is given by \cite{nkfp}
\bea
D_{q \rightarrow \eta^c}(P) = {\cal P}_{q \rightarrow c{\bar c}} \times |<\eta^c(P)|c{\bar c}>|^2
\label{frgp}
\eea
where
\bea
&& {\cal P}_{q\rightarrow c{\bar c}}=N\int d^4z_3 \int d^4z_2 \int d^4z_1 \int d^4z'_3 \int d^4z'_2 \int d^4z'_1\nonumber \\
&& e^{ip_3 \cdot (z_3-z_3')+ip_2 \cdot (z_2-z_2')-ip_1 \cdot (z_1-z_1')}\int d^4x_3 \int d^4x_2 \int d^4x_1  \int d^4x'_3 \int d^4x'_2 \int d^4x'_1 \nonumber \\ && [G_R^c(z_3,x_3)]^{-1}[G_R^c(z_2,x_2)]^{-1}[G_R^q(z_1,x_1)]^{-1}[G_R^c(z'_3,x'_3)]^{-1}[G_R^c(z'_2,x'_2)]^{-1}[G_R^q(z'_1,x'_1)]^{-1}\nonumber \\
&& <0|\psi(x'_1) {\bar \psi}^c(x'_2) \psi^c(x'_3)\psi^c(x_3) {\bar \psi}^c(x_2) \psi(x_1)|0>_R\times [\sum_{\rm spin} |{\bar u}(p_3,\lambda_3)v(p_2,\lambda_2)u(p_1,\lambda_1)|^2]
\label{lsf}
\eea
is the probability for light quark $q$ to fragment to $c{\bar c}$ where $\lambda$ is the helicity of the quark, the suppression of the time ordered product ${\cal T}$ index is understood and \cite{nkhp}
\bea
&&|<\eta^c(P)|c{\bar c}>|^2=\left[\frac{<0|\sum_{\vec z}e^{i{\vec P}\cdot {\vec z}} {\cal B}^{{\eta}^c}({\vec z},\tau){\cal B}^{{\eta}^c}(0)|0>}{e^{ [\frac{<0|\sum_{{\vec z}'}~e^{i{\vec P}\cdot {\vec z}'}{\cal B}^{{\eta}_c}({\vec z}',\tau') [\int d\tau \int d^3z \sum_{q,{\bar q}, g} \partial_j T^{j0}_{\rm Partons}({\vec z},\tau)] {\cal B}^{{\eta}^c}(0)|0>}{<0|\sum_{{\vec z}'}e^{i{\vec P}\cdot {\vec z}'}{\cal B}^{{\eta}^c}({\vec z}',\tau') {\cal B}^{{\eta}^c}(0)|0>}]_{\tau'\rightarrow \infty}}}\right]_{\tau \rightarrow \infty} e^{\tau E^{\eta_c}(P)}\nonumber \\
\label{frgf}
\eea
is the matrix element square for the formation of the hadron $\eta^c$ with momentum $P$ from the charm quark-antiquark pair $c{\bar c}$ where $\tau$ is the Euclidean time (imaginary time).

In eq. (\ref{lsf}) the $N$ is the normalization factor as given by eq. (\ref{nfp}) which arises because in the LSZ reduction formula the creation and annihilation operators are not dimensionless \cite{psp}. The inverse $[G(x',x'')]^{-1}$ of the greens function $G(x',x'')$ in eq. (\ref{lsf}) in the coordinate space is defined by
\bea
\int d^4y' G(z''',y')[G(y',z'')]^{-1}=\delta^{(4)}(z'''-z'').
\label{grft}
\eea
Note that the suppression of color indices in the n-point non-perturbative correlation functions is understood in this paper.

As mentioned above in the LSZ reduction formula the creation and annihilation operators are not dimensionless \cite{psp}. Hence the normalization factor $N$ in eq. (\ref{lsf}) is given by
\bea
N=\frac{1}{<p_1|p_1><p_2,p_3|p_3,p_2>}
\label{nfp}
\eea
where
\bea
&&<p_1|p_1> = \int d^4z'_1 \int d^4z_1 e^{ip_1 \cdot (z'_1-z_1)} \times [\sum_{\rm spin}{\bar u}(p_1,\lambda_1)u(p_1,\lambda_1)]\times \int d^4x'_1 \nonumber \\ && \int d^4y_1 [G_R^q(z'_1,x'_1)]^{-1}[G_R^q(z_1,x_1)]^{-1}<0|{\bar \psi}(x'_1) \psi(x_1)|0>_R
\label{qnfp}
\eea
and
\bea
&&<p_2,p_3|p_3,p_2> = \int d^4z_3 \int d^4z_2 \int d^4z'_3 \int d^4z'_2 e^{ip_3 \cdot (z_3-z'_3)+ip_2 \cdot (z_2-z'_2)} \times [\sum_{\rm spin} {\bar u}(p_3,\lambda_3) u(p_3,\lambda_3)]\nonumber \\
&&[\sum_{\rm spin} v(p_2,\lambda_2) {\bar v}(p_2,\lambda_2)]  \int d^4x_3 \int d^4x_2 \int d^4x'_3 \int d^4x'_2 [G_R^c(z_3,x_3)]^{-1}[G_R^c(z_2,x_2)]^{-1}[G_R^c(z'_3,x'_3)]^{-1}\nonumber \\
&&[G_R^c(z'_2,x'_2)]^{-1}<0|\Psi(x_3) {\bar \Psi}(x_2) \Psi(x'_2) {\bar \Psi}(x'_3)|0>_R.
\label{cnfp}
\eea

The $|<\eta^c(P)|c{\bar c}>|^2$ in eq. (\ref{frgf}) can be calculated by using the lattice QCD method in the Euclidean time \cite{nkhp}. Similarly the probability $ {\cal P}_{q\rightarrow c{\bar c}}$ for the light quark $q$ to fragment to $c{\bar c}$ in eq. (\ref{lsf}) can be calculated by using the lattice QCD method in Euclidean time because the non-perturbative correlation functions $G^c(x',x'')$, $G^q(x',x'')$, $<0|\psi(x'_1) {\bar \psi}^c(x'_2) \psi^c(x'_3)\psi^c(x_3) {\bar \psi}^c(x_2) \psi(c_1)|0>$ in eq. (\ref{lsf}), the $<0|{\bar \psi}(x'_1) \psi(x_1)|0>$ in eq. (\ref{qnfp}) and $<0|\Psi(x_3) {\bar \Psi}(x_2) \Psi(x'_2) {\bar \Psi}(x'_3)|0>$ in eq. (\ref{cnfp}) the can be calculated by using the lattice QCD method in the Euclidean time \cite{nkfp}.

Since the  $|<\eta^c(P)|c{\bar c}>|^2$ in eq. (\ref{frgf}) and the ${\cal P}_{q\rightarrow c{\bar c}}$ in eq. (\ref{lsf}) can be calculated by using the lattice QCD method in Euclidean time one finds that the parton to hadron fragmentation function $D_{q \rightarrow \eta^c}(P)$ in eq. (\ref{frgp}) can be calculated by using lattice QCD method at the zero temperature in the Euclidean time.

\section{ Fragmentation function and LSZ reduction formula in lattice QCD method at finite temperature }

Let us consider the partonic process
\bea
q \rightarrow c{\bar c} +X,~~~~~~~~{\rm in~finite~temperature~QCD}.
\label{qfrp}
\eea
Extending the LSZ reduction formula \cite{nklsp} from the QCD in vacuum \cite{nkfp} to finite temperature QCD we find that the transition amplitude $<c{\bar c}+X|q>$ in finite temperature QCD is given by
\bea
&&<c{\bar c}+X|q> = \int d^4z_3 \int d^4z_2 \int d^4z_1 e^{ip_3 \cdot z_3+ip_2 \cdot z_2-ip_1 \cdot z_1} \int d^4x_3 \int d^4x_2 \int d^4x_1 \nonumber \\ && [G_R^c(z_3,x_3)]^{-1}[G_R^c(z_2,x_2)]^{-1}[G_R^q(z_1,x_1)]^{-1}<X|\psi^c(x_3) {\bar \psi}^c(x_2) \psi(x_1)|in>_R\nonumber \\
&&\times {\bar u}(p_3,\lambda_3)v(p_2,\lambda_2)u(p_1,\lambda_1)
\label{lszp}
\eea
where $|in>$ is the non-perturbative ground state of the QCD at the finite temperature $T$, the $\psi$ is the quark field of the light quark $q$, the $\psi^c$ is the heavy (charm) quark field, $R$ stands for renormalized quantities and the suppression of the time ordered product ${\cal T}$ index is understood. In eq. (\ref{lszp})
\bea
&& G^c(z',z'')
=\frac{1}{Z[0]} \int [d{\bar \psi}_1][d\psi_1][d{\bar \psi}_2][d\psi_2][d{\bar \psi}_3][d\psi_3] [d{\bar \psi}^c][d\psi^c][dA] \nonumber \\
&& \times {\bar \psi}^c(z') \psi^c(z'')~{\rm det}[\frac{\delta H^d}{\delta \omega^c}] \times~{\rm exp}[-\int_0^{\frac{1}{T}} d\tau \int d^3r [-\frac{1}{4} F_{\sigma \mu}^b(\tau,r)F^{\sigma \mu b}(\tau,r) -\frac{1}{2\alpha} [H^d(\tau,r)]^2 +\sum_{f=1}^3 {\bar \psi}_i^f(\tau,r)\nonumber \\
&&[\delta^{ik}(i{\not \partial}-m_f)+gT^b_{ik}\aslash^b(\tau,r)]\psi_k^f(\tau,r)+{\bar \psi}^c_i(\tau,r)[\delta^{ik}(i{\not \partial}-M)+gT^b_{ik}\aslash^b(\tau,r)]\psi_k^c(\tau,r)]],
\label{npp}
\eea
and
\bea
&& G^q(z',z'')
=\frac{1}{Z[0]} \int [d{\bar \psi}_1][d\psi_1][d{\bar \psi}_2][d\psi_2][d{\bar \psi}_3][d\psi_3] [d{\bar \psi}^c][d\psi^c][dA] \nonumber \\
&& \times {\bar \psi}(z') \psi(z'')~{\rm det}[\frac{\delta H^d}{\delta \omega^c}] \times~{\rm exp}[-\int_0^{\frac{1}{T}} d\tau \int d^3r [-\frac{1}{4} F_{\sigma \mu}^b(\tau,r)F^{\sigma \mu b}(\tau,r) -\frac{1}{2\alpha} [H^d(\tau,r)]^2 +\sum_{f=1}^3 {\bar \psi}_i^f(\tau,r)\nonumber \\
&&[\delta^{ik}(i{\not \partial}-m_f)+gT^b_{ik}\aslash^b(\tau,r)]\psi_k^f(\tau,r)+{\bar \psi}^c_i(\tau,r)[\delta^{ik}(i{\not \partial}-M)+gT^b_{ik}\aslash^b(\tau,r)]\psi_k^c(\tau,r)]]
\label{np3p}
\eea
where $Z[0]$ is the generating functional in QCD at the finite temperature given by
\bea
&& Z[0]= \int [d{\bar \psi}_1][d\psi_1][d{\bar \psi}_2][d\psi_2][d{\bar \psi}_3][d\psi_3] [d{\bar \psi}^c][d\psi^c][dA] \nonumber \\
&& \times {\rm det}[\frac{\delta H^d}{\delta \omega^c}] \times~{\rm exp}[-\int_0^{\frac{1}{T}} d\tau \int d^3r [-\frac{1}{4} F_{\sigma \mu}^b(\tau,r)F^{\sigma \mu b}(\tau,r) -\frac{1}{2\alpha} [H^d(\tau,r)]^2 +\sum_{f=1}^3 {\bar \psi}_i^f(\tau,r)\nonumber \\
&&[\delta^{ik}(i{\not \partial}-m_f)+gT^b_{ik}\aslash^b(\tau,r)]\psi_k^f(\tau,r)+{\bar \psi}^c_i(\tau,r)[\delta^{ik}(i{\not \partial}-M)+gT^b_{ik}\aslash^b(\tau,r)]\psi_k^c(\tau,r)]].\nonumber \\
\label{z0tp}
\eea

From eq. (\ref{lszp}) we find that the probability ${\cal P}_{q\rightarrow {\bar c}c}$ for the light quark $q$ to fragment to $ {\bar c}c$ in finite temperature QCD is given by
\bea
&& {\cal P}_{q\rightarrow {\bar c}c}=N_T \sum_X <q|{\bar c}c+X><X+c{\bar c}|q>=N_T\int d^4x_3 \int d^4z_2 \int d^4z_1 \int d^4z'_3 \int d^4z'_2 \int d^4z'_1 \nonumber \\
&&e^{ip_3 \cdot (z_3-z_3')+ip_2 \cdot (z_2-z_2')-ip_1 \cdot (z_1-z_1')}\int d^4x_3 \int d^4x_2 \int d^4x_1  \int d^4x'_3 \int d^4x'_2 \int d^4x'_1 \nonumber \\ && [G_R^c(z_3,x_3)]^{-1}[G_R^c(z_2,x_2)]^{-1}[G_R^q(z_1,x_1)]^{-1}[G_R^c(z'_3,x'_3)]^{-1}[G_R^c(z'_2,x'_2)]^{-1}[G_R^q(z'_1,x'_1)]^{-1}\nonumber \\
&& \sum_X <in|\psi(x'_1) {\bar \psi}^c(x'_2) \psi^c(x'_3)|X><X|\psi^c(x_3) {\bar \psi}^c(x_2) \psi(x_1)|in>_R\times [\sum_{\rm spin} |{\bar u}(p_3,\lambda_3)v(p_2,\lambda_2)u(p_1,\lambda_1)|^2] \nonumber \\
\label{lsz1p}
\eea
which gives
\bea
&& {\cal P}_{q\rightarrow {\bar c}c}=N_T\int d^4z_3 \int d^4z_2 \int d^4z_1 \int d^4z'_3 \int d^4z'_2 \int d^4z'_1 e^{ip_3 \cdot (z_3-z_3')+ip_2 \cdot (z_2-z_2')-ip_1 \cdot (z_1-z_1')}\nonumber \\
&&\int d^4x_3 \int d^4x_2 \int d^4x_1  \int d^4x'_3 \int d^4x'_2 \int d^4x'_1 \nonumber \\ && [G_R^c(z_3,x_3)]^{-1}[G_R^c(z_2,x_2)]^{-1}[G_R^q(z_1,x_1)]^{-1}[G_R^c(z'_3,x'_3)]^{-1}[G_R^c(z'_2,x'_2)]^{-1}[G_R^q(z'_1,x'_1)]^{-1}\nonumber \\
&& <in|\psi(x'_1) {\bar \psi}^c(x'_2) \psi^c(x'_3)\psi^c(x_3) {\bar \psi}^c(x_2) \psi(x_1)|in>_R\times [\sum_{\rm spin} |{\bar u}(p_3,\lambda_3)v(p_2,\lambda_2)u(p_1,\lambda_1)|^2] \nonumber \\
\label{lsz2p}
\eea
where the normalization factor $N_T$ is given by eq. (\ref{nfpt}) and
\bea
&&<in| \psi(x'_1) {\bar \psi}^c(x'_2) \psi^c(x'_3)\psi^c(x_3) {\bar \psi}^c(x_2) \psi(x_1)|in>=\frac{1}{Z[0]} \int [d{\bar \psi}_1][d\psi_1][d{\bar \psi}_2][d\psi_2][d{\bar \psi}_3][d\psi_3] [d{\bar \psi}^c][d\psi^c][dA] \nonumber \\
&& \times \psi(x'_1) {\bar \psi}^c(x'_2) \psi^c(x'_3)\psi^c(x_3) {\bar \psi}^c(x_2) \psi(x_1)\times
{\rm det}[\frac{\delta H^d}{\delta \omega^c}] \times~{\rm exp}[-\int_0^{\frac{1}{T}} d\tau \int d^3r [-\frac{1}{4} F_{\sigma \mu}^b(\tau,r)F^{\sigma \mu b}(\tau,r) \nonumber \\
&&-\frac{1}{2\alpha} [H^d(\tau,r)]^2 +\sum_{f=1}^3 {\bar \psi}_i^f(\tau,r)[\delta^{ik}(i{\not \partial}-m_f)+gT^b_{ik}\aslash^b(\tau,r)]\psi_k^f(\tau,r)+{\bar \psi}^c_i(\tau,r)[\delta^{ik}(i{\not \partial}-M)\nonumber \\
&&+gT^b_{ik}\aslash^b(\tau,r)]\psi_k^c(\tau,r)]].
\label{npea1p}
\eea
In eq. (\ref{lsz2p}) the normalization factor $N_T$ is given by [similar to $N$ in eq. (\ref{nfp})]
\bea
N_T=\frac{1}{<p_1|p_1><p_2,p_3|p_3,p_2>}
\label{nfpt}
\eea
where
\bea
&&<p_1|p_1> = \int d^4z'_1 \int d^4z_1 e^{ip_1 \cdot (z'_1-z_1)} \times [\sum_{\rm spin}{\bar u}(p_1,\lambda_1)u(p_1,\lambda_1)]\times \int d^4x'_1 \nonumber \\ && \int d^4y_1 [G_R^q(z'_1,x'_1)]^{-1}[G_R^q(z_1,x_1)]^{-1}<in|{\bar \psi}(x'_1) \psi(x_1)|in>_R
\label{qnfpt}
\eea
and
\bea
&&<p_2,p_3|p_3,p_2> = \int d^4z_3 \int d^4z_2 \int d^4z'_3 \int d^4z'_2 e^{ip_3 \cdot (z_3-z'_3)+ip_2 \cdot (z_2-z'_2)} \times [\sum_{\rm spin} {\bar u}(p_3,\lambda_3) u(p_3,\lambda_3)]\nonumber \\
&&[\sum_{\rm spin} v(p_2,\lambda_2) {\bar v}(p_2,\lambda_2)]  \int d^4x_3 \int d^4x_2 \int d^4x'_3 \int d^4x'_2 [G_R^c(z_3,x_3)]^{-1}[G_R^c(z_2,x_2)]^{-1}[G_R^c(z'_3,x'_3)]^{-1}\nonumber \\
&&[G_R^c(z'_2,x'_2)]^{-1}<in|\Psi(x_3) {\bar \Psi}(x_2) \Psi(x'_2) {\bar \Psi}(x'_3)|in>_R.
\label{cnfpt}
\eea
Since the non-perturbative n-point correlation functions $G^c(z'',z''')$, $G^q(z'',z''')$, $<in| \psi(x'_1) {\bar \psi}^c(x'_2) \psi^c(x'_3)\psi^c(x_3) {\bar \psi}^c(x_2) \psi(x_1)|in>$, $<in|{\bar \psi}(x'_1) \psi(x_1)|in>$ and $<in|\Psi(x_3) {\bar \Psi}(x_2) \Psi(x'_2) {\bar \Psi}(x'_3)|in>$ in eqs. (\ref{npp}), (\ref{np3p}), (\ref{npea1p}), (\ref{qnfpt}) and (\ref{cnfpt}) respectively in the finite temperature QCD can be calculated by using the lattice QCD method at the finite temperature in the Euclidean time one finds that the probability ${\cal P}_{q\rightarrow {\bar c}c}$ for the light quark $q$ to fragment to $ {\bar c}c$ in the finite temperature QCD in eq. (\ref{lsz2p}) can be calculated by using the lattice QCD method at the finite temperature in the Euclidean time.

\section{ Parton To Hadron Fragmentation function from the quark-gluon plasma by using lattice QCD method at the finite temperature}

Extending eq. (\ref{frgf}) from zero temperature QCD to finite temperature QCD we find \cite{nkftp}
\bea
&&|<\eta^c(P)|c{\bar c}>|^2=\left[\frac{\sum_{r'''} <in|e^{-\tau'''H}{\hat {\cal B}}^{\eta^c}(\tau''',r''') {\hat {\cal B}}^{\eta^c}(0)|in>}{e^{[\frac{<0| \sum_{r''}{\hat {\cal B}}^{\eta^c}(\tau'',r'') \sum_{q,{\bar q},g} \int d\tau''' \int d^3r''' \partial_j  T^{j 0}_{q{\bar q}g}(\tau''',r'''){\hat {\cal B}}^{\eta^c}(0) |0>}{<0| \sum_{r''}{\hat {\cal B}}^{\eta^c}(\tau'',r'') {\hat {\cal B}}^{\eta^c}(0) |0>}]_{\tau''\rightarrow \infty}}}\right]_{\tau'''\rightarrow \infty}\times ~e^{\tau'''M^H}\nonumber \\
\label{frgfp}
\eea
where $|<\eta^c(P)|c{\bar c}>|^2$ is the matrix element square for the formation of the heavy quarkonium $\eta^c$ from the $c{\bar c}$ by using the lattice QCD method at the finite temperature in the Euclidean time and
\bea
&&<in|{\hat {\cal B}}^{\eta^c}(\tau''',r''') {\hat {\cal B}}^{\eta^c}(0)|in>=\frac{1}{Z[0]} \int [d{\bar \psi}_1][d\psi_1][d{\bar \psi}_2][d\psi_2][d{\bar \psi}_3][d\psi_3] [d{\bar \psi}^c][d\psi^c][dA] \nonumber \\
&& \times {\hat {\cal B}}^{\eta^c}(\tau''',r''') {\hat {\cal B}}^{\eta^c}(0)\times
{\rm det}[\frac{\delta H^d}{\delta \omega^c}] \times~{\rm exp}[-\int_0^{\frac{1}{T}} d\tau \int d^3r [-\frac{1}{4} F_{\sigma \mu}^b(\tau,r)F^{\sigma \mu b}(\tau,r) \nonumber \\
&&-\frac{1}{2\alpha} [H^d(\tau,r)]^2 +\sum_{f=1}^3 {\bar \psi}_i^f(\tau,r)[\delta^{ik}(i{\not \partial}-m_f)+gT^b_{ik}\aslash^b(\tau,r)]\psi_k^f(\tau,r)+{\bar \psi}^c_i(\tau,r)[\delta^{ik}(i{\not \partial}-M)\nonumber \\
&&+gT^b_{ik}\aslash^b(\tau,r)]\psi_k^c(\tau,r)]]
\label{etmi}
\eea
which can be calculated by using the lattice QCD method at the finite temperature in the Euclidean time.

From eqs. (\ref{lsz2p}) and (\ref{frgfp}) we find that the fragmentation function $D_{q\rightarrow \eta^c}$ for the light quark $q$ from the quark-gluon plasma at the finite temperature to fragment to the heavy quarkonium $\eta^c$ is given by
\bea
D_{q\rightarrow \eta^c} (P)={\cal P}_{q\rightarrow {\bar c}c} \times |<\eta^c(P)|c{\bar c}>|^2
\label{frfnp}
\eea
where the probability ${\cal P}_{q\rightarrow {\bar c}c}$ for the light quark $q$ to fragment to $ {\bar c}c$ in finite temperature QCD is given by eq. (\ref{lsz2p})
and the $|<\eta^c(P)|c{\bar c}>|^2$ is the matrix element square for the formation of the heavy quarkonium $\eta^c$ from the $c{\bar c}$ as given by eq. (\ref{frgfp}).

Since the probability ${\cal P}_{q\rightarrow {\bar c}c}$ for the light quark $q$ to fragment to $ {\bar c}c$ in finite temperature QCD in eq. (\ref{lsz2p}) can be calculated by using the lattice QCD method at the finite temperature in the Euclidean time and the matrix element square $|<\eta^c(P)|c{\bar c}>|^2$ for the formation of the heavy quarkonium $\eta^c$ from the $c{\bar c}$ in eq. (\ref{frgfp}) can be calculated by using the lattice QCD method at the finite temperature in the Euclidean time we find that the fragmentation function $D_{q\rightarrow \eta^c}(P)$ in eq. (\ref{frfnp}) for the light quark $q$ from the quark-gluon plasma at the finite temperature $T$ to fragment to $\eta^c$ with momentum $P$ can be calculated by using the lattice QCD method at finite temperature in the Euclidean time.

\section{Conclusions}
Recently we have shown that it is possible to study the parton to hadron fragmentation function by using the lattice QCD method at the zero temperature. In this paper we have extended this to the finite temperature QCD and have studied the parton to hadron fragmentation function from the quark-gluon plasma by using the lattice QCD method at the finite temperature.

\end{document}